# Approximate Analytical Solutions of Power Flow Equations Based on Multi-Dimensional Holomorphic Embedding Method

Chengxi Liu *Member, IEEE*, Bin Wang, *Student Member*, *IEEE*, Xin Xu, *Student Member*, *IEEE*, Kai Sun, *Senior Member IEEE*, Claus Leth Bak, *Senior Member IEEE*

***Abstract*—It is well known that closed-form analytical solutions for AC power flow equations do not exist in general. This paper proposes a multi-dimensional holomorphic embedding method (MDHEM) to obtain an explicit approximate analytical AC power-flow solution by finding a physical germ solution and arbitrarily embedding each power, each load or groups of loads with respective scales. Based on the MDHEM, the complete approximate analytical solutions to the power flow equations in the high-dimensional space become achievable, since the voltage vector of each bus can be explicitly expressed by a convergent multivariate power series of all the loads. Unlike the traditional iterative methods for power flow calculation and inaccurate sensitivity analysis method for voltage control, the algebraic variables of a power system in all operating conditions can be prepared offline and evaluated online by only plugging in the values of any operating conditions into the scales of the non-linear multivariate power series. Case studies implemented on the 4-bus test system and the IEEE 14-bus standard system confirm the effectiveness of the proposed method.**

## I. Introduction

FAST growing electricity markets and relatively slow upgrades on transmission infrastructure have made many power systems occasionally operated closer to their voltage stability limits. However, many blackouts, such as the western North America blackout in July, 1996 [1], and Indian power system blackout in July, 2012 [2] resulted from voltage collapses have led to enormous societal and financial losses. To prevent voltage collapses, many utilities have deployed different assessment and early awareness systems on the transmission grids. However, the accurate situational awareness and the corresponding preventive control will only be achieved when the control center have the fast, reliable and accurate tools to directly process these measured data preferably in real-time.

Power flow analysis on a power system is necessary for utilities to provide the voltage stability assessment, situational awareness and preventive control. Theoretically, power-flow equations (PFEs) of power systems are a set of non-linear algebraic equations reduced from an enormous number of detailed differential equations by neglecting the fast electromagnetic dynamics. Traditionally, to solve AC PFEs, many iterative numerical methods have been adopted by commercialized power system software, including the Gauss-Seidel method, Newton-Raphson method and fast decoupled method. A major concern on these methods is that the numerical divergence of their iterations is often interpreted as the happening of voltage collapse but, theoretically speaking, does not necessarily indicate the non-existence of a power flow solution. Also, there is a probability for these numerical methods to converge to the non-physically existing ghost solutions [3]. These concerns influence the performances of these iterative, numerical methods in real-time applications.

The holomorphic embedding power flow method (HELM) was firstly proposed by A. Trias in [3]-[5], which is a non-iterative method to analyze the power flows. The basic idea of HELM is to design a holomorphic function and adopt its analytical continuation in the complex plane to find the solution of the power flow equations as a power series form about an embedded complex variable. Recently, many derivative algorithms and applications based on HELM have developed [6]-[9], such as the HELM with non-linear static load models [10], the HELM used in AC/DC power systems [11], using HELM to find the unstable equilibrium points [12], [13], network reduction [14], the analysis of saddle-node bifurcation [15], [16] and the applications of real-time voltage stability assessment [17].

In this paper, a novel multi-dimensional holomorphic embedding method (MDHEM) is proposed to obtain the approximate analytical solution to PFEs, by finding a germ solution in the space of physical solutions and arbitrarily embedding each load or groups of loads of respective scales. Therefore, the approximate analytical solutions to the PFEs in the high-dimensional solution space become achievable. Based on the proposed MDHEM, the algebraic variables of a power network in all operating conditions can be prepared offline and evaluated online by only plugging in the operating values into the scales of multivariate power series.

The rest of this paper is organized as follows. Section II introduces the conventional HELM. Section III describes the details of the proposed MDHEM, including the physical germ

This work was supported in part by *NSF CURENT Engineering Research Center* and *NSF grant ECCS-1610025*.

C. Liu, B. Wang, X. Xu and K. Sun are with Department of EECS, University of Tennessee, Knoxville, TN, USA (email: cliu48@utk.edu, bwang@utk.edu, xxu30@vols.utk.edu, kaisun@utk.edu,)

C. L. Bak is with Department of Energy, Aalborg University, Aalborg, Denmark (email: clb@et.aau.dk)

solution, the derivation process of the MDHEM with and without PV buses, the transformation procedure of bus types considering reactive power limits of generator buses, the analysis of computational burden and the details of multi-dimensional discrete convolution that is used in the algorithms of the MDHEM. Section IV uses a 4-bus system and the 14-bus standard system to verify the effectiveness of this new algorithm and to evaluate the accuracy of the obtained analytical solutions up to a specific order. Finally, conclusions are drawn in Section V.

## II. Conventional Holomorphic Embedding Load Flow Method

In context of complex analysis, holomorphic functions are complex functions, defined on an open subset of the complex plane, that are differentiable in the neighborhood of every point in its domain. Define a complex function whose domain and range are subsets of the complex plane,

$$f(z)=f(x+iy)=u(x,y)+i\cdot v(x,y) \tag{1}$$

where $x$ and $y$ are real variables and $u(x, y)$ and $v(x, y)$ are real-valued functions. The derivative of function $f$ at $z_0$ is

$$f^{\prime}(z_0)=\lim_{z\to z_0}\frac{f(z)-f(z_0)}{z-z_0}, z\in\mathbf{C} \tag{2}$$

An important property that characterizes holomorphic functions is the relationship between the partial derivatives of their real and imaginary parts, known as Cauchy-Riemann condition, defined in (3). However, based on Looman-Menchoff theorem, functions satisfying Cauchy-Riemann conditions are not necessarily holomorphic, unless the continuity is met [18].

$$i\frac{\partial f}{\partial x}=\frac{\partial f}{\partial y} \tag{3}$$

The HELM is founded on the theory of complex analysis, whose main advantages are its explicit, non-iterative nature, and it mathematically guarantees the convergence to the upper operative branch, i.e. white branch, for multi-value power flow calculation. Additionally, assisted by Pade approximants, it is able to sufficiently and necessarily indicate the condition of voltage collapse when the solution does not exist.

For power flow calculation of a power grid having PV buses, the HELM decomposes the admittance matrix $Y_{ik}$ to a series admittance part $Y_{ik,tr}$ and a shunt admittance part $Y_{i,sh}$. The advantage of this process mainly lies on the simplification of the germ solution that has all bus voltages equal directly 1 $\angle 0^{\circ}$ for the no-load, no-generation and no-shunt condition. See the embedding method of PFEs with the HELM for PQ, PV and slack buses respectively in Table I. The germ solution can be obtained by plugging $s$=0 into expressions on the third column of the table, while the final solution of PFEs can be achieved with $s$=1. Thus, under this circumstance, if the original implicit PFEs regarding voltage vectors can be transformed to the explicit form of a power series, like (4), the final solution can be obtained by plugging $s$=1 into the power series, if only $s$=1 is in the convergence region (refer to [8] for more details).

$$V(s)=\sum_{n=0}^{\infty}V[n]s^n \tag{4}$$

There are several other methods embedding the complex value to solve the original PFEs [9]-[12]. Nevertheless, the common idea of the solving process is to express the nonlinear embedded variables to a Taylor series, e.g. $V(s)$ and $Q(s)$, and then equate both sides of the complex-valued equations to find the coefficients of the Taylor series. Theoretically, similar to the mathematical induction method, the coefficients of the Taylor series can be calculated term by term, under the precondition that Taylor series can approximate to the complex-valued nonlinear holomorphic embedded function.

TABLE I. The embedding of PFE for PQ, PV and slack buses with HELM

| Type | PFE | Holomorphic Embedding Method |
|---|---|---|
| **SL** | $V_i(s)=V_i^{SL}$ | $V_i(s)=1+(V_i^{SL}-1)s$ |
| **PQ** | $\sum_{k=1}^{N}Y_{ik}V_k(s)=S_i^*/V_i^*$ | $\sum_{k=1}^{N}Y_{ik,tr}V_k(s)=\frac{sS_i^*}{V_i^*(s^*)}-sY_{i,sh}V_i(s)$ |
| **PV** | $P_i=\mathrm{Re}\left(V_i\sum_{k=1}^{N}Y_{ik}^*V_k^*\right)$<br>$\lvert V_i\rvert=V_i^{sp}$ | $\sum_{k=1}^{N}Y_{ik,tr}V_k(s)=\frac{sP_i-jQ_i(s)}{V_i^*(s^*)}-sY_{i,sh}V_i(s)$<br>$V_i(s)\cdot V_i^*(s^*)=1+\left(\lvert V_i^{sp}\rvert^2-1\right)s$ |

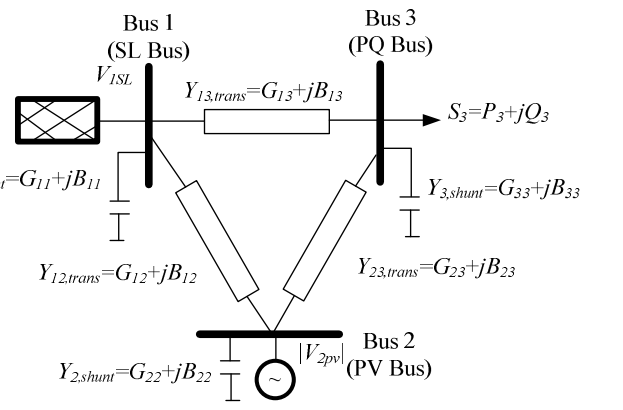

Fig. 1. One-line diagram of the demonstrative 3-bus system.

For the sake of simplification, let us take a simple three bus system for example, as shown in Fig. 1, which has one slack bus, one PQ bus and one PV bus. The equation of PV bus, i.e. Column 3 Row 3 of Table I, can be equated as (5) and the $n$th term can be calculated as (6) by equating both sides of (5) with the same order of $s^n$.

$$\sum_{k=1}^{N}Y_{iktrans}\left(V_k[0]+V_k[1]s+\cdots\right) \tag{5}$$
$$=\left(sP_i-j\left(Q_i[0]+Q_i[1]s+\cdots\right)\right)\cdot\left(W_i^*[0]+W_i^*[1]s+\cdots\right)-sY_{ishunt}\left(V_i[0]+V_i[1]s+\cdots\right)$$

$$\sum_{k=1}^{N}Y_{iktrans}V_k[n]=P_i\cdot W_i^*[n-1]-j\left(Q_i[n]+\overset{n-1}{\underset{1}{Conv}}(Q*W^*)\right)-Y_{ishunt}V_i[n-1] \tag{6}$$

where $W^*(s)$ is defined as the reciprocal power series of $V^*(s^*)$.

$$W^*(s)=\frac{1}{V^*(s^*)}=W_i^*[0]+W_i^*[1]s+W_i^*[2]s^2+\cdots \tag{7}$$

Therefore, given the germ solution, i.e. $V$[0]=1 for this embedding method, the coefficients of $W^*[n]$ can be calculated by a single-dimensional convolution between $W^*(s)$ and $V^*(s^*)$.

$$\begin{cases}W^*[0]=1/V_k^*[0] & \text{for } n=0\\ W^*[n]=-\sum_{\tau=0}^{n-1}W^*[\tau]V^*[n-\tau]\Big/V_k^*[0] & \text{for } n\geq 1\end{cases} \tag{8}$$

$$\begin{bmatrix}1&&&&&\\&1&&&&\\G_{21}&-B_{21}&0&-B_{22}&G_{23}&-B_{23}\\B_{21}&G_{21}&1&G_{22}&B_{23}&G_{23}\\G_{31}&-B_{31}&0&-B_{32}&G_{33}&-B_{33}\\B_{31}&G_{31}&0&G_{32}&B_{33}&G_{33}\end{bmatrix}\begin{bmatrix}V_{1re}[n]\\V_{1im}[n]\\Q_2[n]\\V_{2im}[n]\\V_{3re}[n]\\V_{3im}[n]\end{bmatrix}=\begin{bmatrix}\delta_{n0}+\delta_{n1}\left(V_1^{SL}-1\right)\\0\\\mathrm{Re}\left(P_2W_2^*[n-1]-j\overset{n-1}{\underset{1}{Conv}}(Q_2*W_2^*)-Y_{2,sh}V_2[n-1]\right)\\\mathrm{Im}\left(P_2W_2^*[n-1]-j\overset{n-1}{\underset{1}{Conv}}(Q_2*W_2^*)-Y_{2,sh}V_2[n-1]\right)\\\mathrm{Re}\left(S_3^*W_3^*[n-1]-Y_{3,sh}V_3[n-1]\right)\\\mathrm{Im}\left(S_3^*W_3^*[n-1]-Y_{3,sh}V_3[n-1]\right)\end{bmatrix}-\begin{bmatrix}0\\0\\G_{22}\\B_{22}\\G_{32}\\B_{32}\end{bmatrix}V_{2re}[n] \tag{9}$$

Similarly, the coefficients of $Q(s)$ can also be solved by single-dimensional convolution between $W^*(s)$ and $Q(s)$. Finally, the holomorphic embedded PFE of three bus system is separated into the real part and imaginary part, and expressed as the mathematical induction form, where the $n$th term on the left hand side is dependent on terms 0 to $(n\text{-}1)^{th}$ on the right hand side of the equation (9). In (9), $V_{2re}[n]$ is the real part of PV bus dependent on the $1$-$(n\text{-}1)$th order of $V_2$.

$$V_{2re}[n] = \delta_{n0} + \delta_{n1}\frac{\left(V_2^{sp}\right)^2 - 1}{2} - \frac{1}{2}\sum_{\tau=1}^{n-1} V_2[\tau]V_2^*[n-\tau] \tag{10}$$

## III. Multi-Dimensional Holomorphic Embedding Method

Theoretically, given enough precision digits in numeric arithmetic, a conventional HEM, e.g. the HELM, can find the power flow solution at one required operating condition with high accuracy using a number of terms in Taylor series. However, the main drawback is that it cannot give the expression of the power flow solution at any operating condition, so the explicit expression for the *whole solution space* of the PFEs is impossible. The first precondition of this analytical expression is to find a *physical germ solution*, which serves as an original point in the solution space. Secondly, extend from the physical germ solution using the MDHEM by endowing the embedded variables with physical meanings, i.e. the loading scales to control the loading levels of load buses. We may adopt a scale either for each load or for each group of loads. Details of the calculation method are introduced in the following.

### A. Physical Germ Solution

The proposed physical germ solution is an original point of the solution at an operating condition with physical meaning. In this paper, the physical germ solution is the operating condition with no-load no-generation for load buses (PQ buses), while active power is specified and reactive power is injected to control the voltage magnitude to a specified value at each generator bus (PV bus). Although other physical germ solutions can be defined, there are advantages of this physical germ solution. First, this proposed physical germ solution is easy to obtain and it is origin of the solution space that considers the power generation and voltage control at PV buses and no-load at PQ buses. Second, with the correct embedding method, the embedded variables in the multivariate complex function can easily be ensured to be the loading scales used to extend the operating condition to the whole multi-dimensional space.

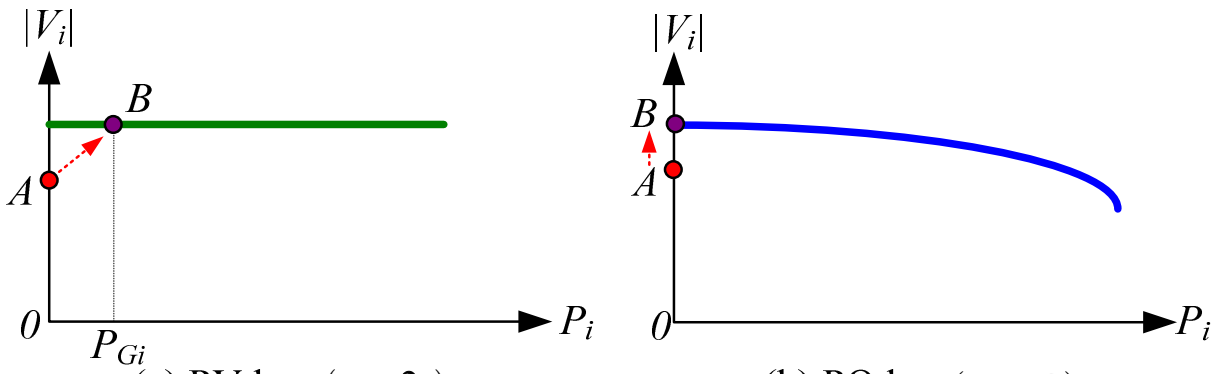

(a) PV bus ($\forall i \in \mathcal{V}$) (b) PQ bus ($\forall i \in \mathcal{P}$)

Fig. 2. The procedure of finding the physical germ solution.

The procedure of finding this physical germ solution consists of two steps, illustrated in Fig. 2. The first step is to obtain the initial condition, i.e. point $A$, under which the slack bus is the only voltage source propagating its voltage to every bus of the network, while PV and PQ buses have zero injection to the grid. Equations (11)-(13) define the first step from the physical germ solution, where slack buses, PV buses and PQ buses are denoted as sets of $\mathcal{S}$, $\mathcal{V}$ and $\mathcal{P}$ respectively.

$$V_{gi}[0] = V_i^{SL}, \quad \forall i \in \mathcal{S} \tag{11}$$

$$\sum_{k=1}^{N} Y_{ik}V_{gi}[0] = 0, \forall i \in \mathcal{P} \tag{12}$$

$$\sum_{k=1}^{N} Y_{ik}V_{gi}[0] = 0, \forall i \in \mathcal{V} \tag{13}$$

Then the second step is to embed a series of reactive power $Q_g(s)$ for PV buses to control their voltage magnitudes to the specified voltage $|V^{sp}|$, i.e. point $B$. Meanwhile, the active power of each PV bus is fixed at the base value of the original condition, i.e. $P_{Gi}$ in Fig. 2(a).

TABLE II shows the two-step process of finding the physical germ solution for different bus types. Given the $N$-bus network with $s$ slack buses, $p$ PQ buses and $v$ PV buses, it is pertinent to mention that the dimension of the matrix equation is extended from $2s$+$2p$+$2v$ in (9) to $2s$+$2p$+$5v$, since a complex valued series $W(s)$ and a real valued series $Q(s)$ are added to the equation for each PV bus. An example of the demonstrative 3-bus system to find the physical germ solution is introduced in detail in Appendix-A.

TABLE II. Physical germ solution of PFE with the MDHEM

| Type | 1st Step (*n* = 0) | 2nd Step (*n*>= 1) |
|---|---|---|
| **SL** | $V_{gi}[0] = V_i^{SL}$ | $V_{gi}(s) = 0$ |
| **PQ** | $\sum_{k=1}^{N} Y_{ik}V_{gi}[0] = 0$ | $\sum_{k=1}^{N} Y_{ik}V_{gi}(s) = 0$ |
| **PV** | $\sum_{k=1}^{N} Y_{ik}V_{gi}[0] = 0$ | $\sum_{k=1}^{N} Y_{ik}V_{gi}(s) = \frac{sP_{gi} - jQ_{gi}(s)}{V_{gi}^*(s^*)}$ <br> $V_{gi}(s)V_{gi}^*(s^*) = \lvert V_{STi}\rvert^2 + \left(\lvert V_{gi}^{sp}\rvert^2 - \lvert V_{STi}\rvert^2\right)s$ |

### B. From Single-Dimension to Multi-Dimension

As mentioned earlier, the conventional single-dimensional HEM has one major drawback, which scales all loads in the system uniformly at the same rate, since only one $s$ is embedded. Loads cannot decrease, remain either static or grow at separate rates, and the power factor of each load is fixed. Therefore, it is impossible to find the solution for all the operating conditions in the solution space of PFEs. The MDHEM is proposed here to obtain complete solutions to the PFEs in the high-dimensional space. The analytical expression is derived from the physical germ solution by endowing multiple embedded variables, i.e. $s_1$, $s_2$, …, $s_n$, with different physical meanings. Each loading scale $s_i$ can control one active power or reactive power at a load bus, or control one load or a group of loads. The scaling method can be user defined. As the special case of the MDHEM, the single dimensional HEM can be adopted perform continuous power flow in an extremely fast way. Conversely, as another special case of MDHEM in which each scale only control one active or reactive power, the analytical expression of the whole solution space can be obtained.

For the sake of clarity, the MDHEM will be firstly exposed in the case without PV bus. The embedding can be done by scaling each active or reactive power separately, so each active power or reactive power is multiplied by respective $s$ in the PFE of PQ buses. A $D$-dimensional HEM is defined in (14),

$$\sum_{k=1}^{N} Y_{ik} V_k(s_1, s_2, \cdots, s_D) = \frac{s_{ip} P_i - j s_{iq} Q_i}{V_i^*(s_1^*, s_2^*, \cdots, s_D^*)}, \quad i \in \mathcal{P} \tag{14}$$

where $V_i(s_1, s_2, \ldots, s_D)$ is a multivariate power series (15) for the voltage vector at bus $i$, whose reciprocal is defined by a new multivariate power series $W_i(s_1, s_2, \ldots, s_D)$.

$$\begin{aligned} V_i(s_1, s_2, \cdots, s_D) &= \sum_{n_D=0}^{\infty} \cdots \sum_{n_2=0}^{\infty} \sum_{n_1=0}^{\infty} V[n_1, n_2, \cdots, n_D] s_1^{n_1} s_2^{n_2} \cdots s_D^{n_D} \\ &= V_i[\underbrace{0, 0, \cdots, 0}_{D-\text{dimension}}] + V_i[1, 0, \cdots, 0] s_1 + V_i[0, 1, \cdots, 0] s_2 + \cdots \\ &\quad + V_i[2, 0, \cdots, 0] s_1^2 + V_i[1, 1, \cdots, 0] s_1 s_2 + V_i[0, 2 \cdots, 0] s_2^2 + \cdots \end{aligned} \tag{15}$$

The following equation is thus obtained from (14) and (15), where power series appear on both sides.

$$\begin{aligned} &\sum_{k=1}^{N} Y_{ik} \left( V_k[0, 0, \cdots 0] + V_k[1, 0, \cdots, 0] s_1 + V_k[0, 1, \cdots, 0] s_2 + \cdots \right) \\ &= \left( s_{ip} P_i - j s_{iq} Q_i \right) \bullet \left( W_i^*[0, 0, \cdots 0] + W_i^*[1, 0, \cdots, 0] s_1 + W_i^*[0, 1, \cdots, 0] s_2 + \cdots \right) \end{aligned} \tag{16}$$

The summation of $n_1$, $n_2$, …, $n_D$, denoted as $M$, is the order of recursion, so $V_k[0,0,\ldots,0]$ is the physical germ solution for $M$=0. Equate both sides of (16) and then extend the matrix equation for the $M$th order as (17).

$$\begin{aligned} &[Y_{ik}]_{N \times N} \begin{bmatrix} V_1[M, 0, \cdots 0] & V_1[M-1, 1, 0, \cdots 0] & \cdots & V_1[0, 0, \cdots M] \\ V_2[M, 0, \cdots 0] & V_2[M-1, 1, 0, \cdots 0] & \cdots & V_2[0, 0, \cdots M] \\ \vdots & \vdots & \ddots & \vdots \\ V_N[M, 0, \cdots 0] & V_N[M-1, 1, 0, \cdots 0] & \cdots & V_N[0, 0, \cdots M] \end{bmatrix}_{N \times N_{col}} \\ &= \begin{bmatrix} 0 & 0 & 0 & 0 \\ \vdots & \ddots & \ddots & \vdots \\ \vdots & P_i W_i^*[\cdots, n_{ip} - 1, \cdots] & -j Q_i W_i^*[\cdots, n_{iq} - 1, \cdots] & \vdots \\ \vdots & \ddots & \ddots & \vdots \end{bmatrix}_{N \times N_{col}} \end{aligned} \tag{17}$$

The equation is no longer a fixed $N$×1 matrix equation unlike that in the single dimensional HEM. The number of columns, denoted by $N_{col}$, is a $D$-polytope number expanding with the increase of order $M$. For $D$-dimensional HEM at the $M$th order,

$$N_{col} = \frac{\prod_{i=M+1}^{M+D-1} i}{(D-1)!} = \frac{(M+D-1)!}{M!(D-1)!}. \tag{18}$$

Take a 2-$D$ HEM for example. A 2-bus system with one slack bus and one PQ bus is demonstrated. $s_1$ and $s_2$ are selected to scale the active and reactive powers of the PQ bus, respectively. The $M$th recursion of the matrix calculation is

$$\begin{aligned} &\begin{bmatrix} 1 & 0 & 0 & 0 \\ 0 & 1 & 0 & 0 \\ G_{21} & -B_{21} & G_{22} & -B_{22} \\ B_{21} & G_{21} & G_{21} & G_{22} \end{bmatrix}_{4 \times 4} \begin{bmatrix} V_{1re}[M,0] & V_{1re}[M-1,1] & \cdots & V_{1re}[n_1,n_2] & \cdots & V_{1re}[0,M] \\ V_{1im}[M,0] & V_{1im}[M-1,1] & \cdots & V_{1im}[n_1,n_2] & \cdots & V_{1im}[0,M] \\ V_{2re}[M,0] & V_{2re}[M-1,1] & \cdots & V_{2re}[n_1,n_2] & \cdots & V_{2re}[0,M] \\ V_{2re}[M,0] & V_{2im}[M-1,1] & \cdots & V_{2im}[n_1,n_2] & \cdots & V_{2im}[0,M] \end{bmatrix}_{4 \times (M+1)} \\ &= \begin{bmatrix} 0 & \cdots & 0 & \cdots & 0 \\ 0 & \cdots & 0 & \cdots & 0 \\ \mathrm{Re}\left(P_2 W_2^*[M-1,0]\right) & \cdots & \mathrm{Re}\left(P_2 W_2^*[n_1-1,n_2] - jQ_2 W_2^*[n_1,n_2-1]\right) & \cdots & \mathrm{Re}\left(-jQ_2 W_2^*[0,M-1]\right) \\ \mathrm{Im}\left(P_2 W_2^*[M-1,0]\right) & \cdots & \mathrm{Im}\left(P_2 W_2^*[n_1-1,n_2] - jQ_2 W_2^*[n_1,n_2-1]\right) & \cdots & \mathrm{Im}\left(-jQ_2 W_2^*[0,M-1]\right) \end{bmatrix}_{4 \times (M+1)} \end{aligned} \tag{19}$$

where the 2-$D$ discrete convolution between $W(s_1,s_2)$ and $V(s_1,s_2)$ from 0 to $M$-1 is equal to 1. Therefore, the $(M\text{-}1)^{\text{th}}$ order of $W[n_1,n_2]$ can be calculated from the already obtained terms 0 to $M$-1 of $V[n_1,n_2]$. The matrix equation is separated to real and imaginary parts to be in consistent with the network with PV buses. The details of multi-dimensional discrete convolution used to calculate $W[n_1,n_2]$ are described below.

## C. Multi-Dimensional Discrete Convolution

For the revolution from single-dimensional HEM to multi-dimensional HEM, the single-dimensional discrete convolution on the right hand side of (9) also evolves to the corresponding multi-dimensional discrete convolution in (19).

Multi-dimensional discrete convolution refers to the rolling multiplication operation between two functions (e.g. $f$ and $g$) on an $n$-dimensional lattice that produces a third function, also of $n$-dimensions. Generally, an asterisk is used to represent the convolution operation. The number of dimensions in the given operation is reflected in the number of asterisk. Taking the $D$-dimensional HEM for example, a $D$-dimensional convolution would be written in (20) and can be computed via (21).

$$y(n_1, n_2, \cdots n_D) = f(n_1, n_2, \cdots n_D) \overset{D}{*\cdots*} g(n_1, n_2, \cdots n_D) \tag{20}$$

$$= \sum_{\tau_1=-\infty}^{\infty} \sum_{\tau_2=-\infty}^{\infty} \cdots \sum_{\tau_D=-\infty}^{\infty} f(n_1 - \tau_1, n_2 - \tau_2, \cdots, n_D - \tau_D) \cdot g(\tau_1, \tau_2, \cdots, \tau_D) \tag{21}$$

Different from the conventional multi-dimensional convolution which takes the terms from negative infinity to positive infinity, the $D$-dimensional convolution used in the HEM truncates the part from $1^{\text{st}}$ term to $(n\text{-}1)^{\text{th}}$ term. Therefore, the convolutions of $W[n_1,n_2]$ on the right hand side of (19) are calculated by the following multiplication of two $n$-dimensional arrays. Fig. 3 vividly illustrates the 1-, 2- and 3-dimensional discrete convolution of order from term 0 to term 2, in which the blue and red lattices move and overlap with each other. The convolution is the summation of the multiplication between geometrically super-positioned red lattices and blue lattices. The discrete convolution for higher dimensions is similar but in a space of higher dimensions.

$$\begin{aligned} &\underset{k_i=1}{\overset{k_i=n_i-1}{Conv}}(W * V)[k_1, k_2, \cdots, k_D] \\ &= \sum_{\tau_1=1}^{n_1-1} \sum_{\tau_2=1}^{n_2-1} \cdots \sum_{\tau_D=1}^{n_D-1} W[k_1 - \tau_1, k_2 - \tau_2, \cdots, k_D - \tau_D] \times V[\tau_1, \tau_2, \cdots, \tau_D] \end{aligned} \tag{22}$$

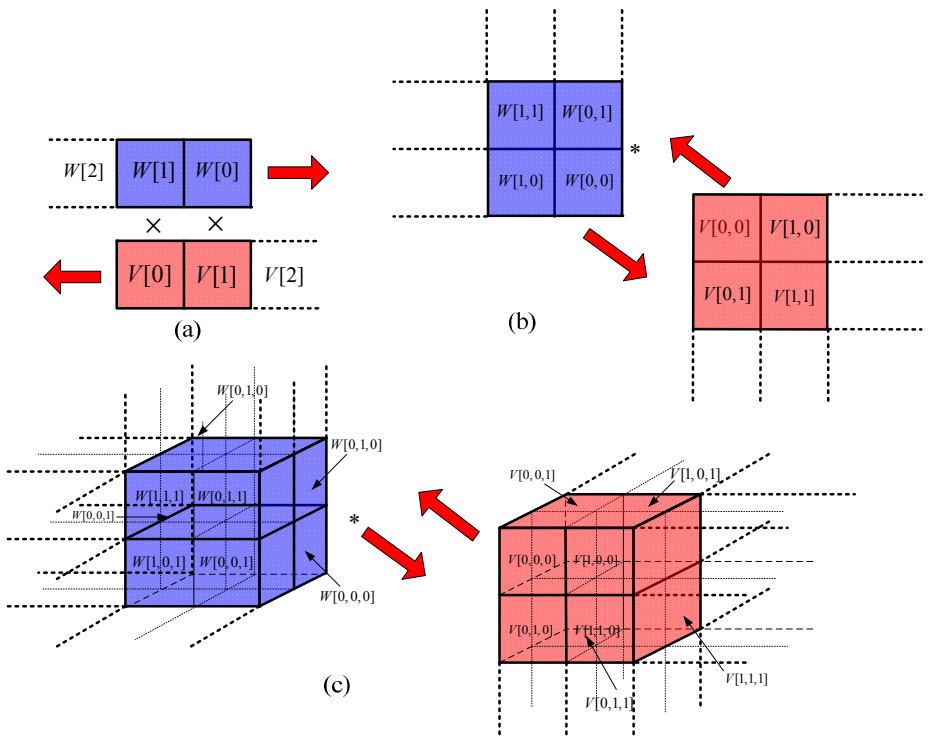

Fig. 3. The illustration of $W*V(2,2,\ldots,2)$ (a) 1-dimensional (b) 2-dimensional (c) 3-dimensioanl discrete convolution.

## D. Multi-Dimensional HEM with PV buses

For a $N$-bus network with $v$ PV buses, the complex PFE for PV buses is (23), with reactive power $Q(s_i)$ also represented as a form of multivariate power series with only real coefficients.

$$\sum_{k=1}^{N} Y_{ik} V_k(s_1, s_2, \cdots, s_D) = \frac{P_i - jQ_i(s_1, s_2, \cdots, s_D)}{V_i^*(s_1^*, s_2^*, \cdots, s_D^*)}, \quad i \in \mathcal{V} \tag{23}$$

The following equation is thus obtained from (23) where multivariate power series appears on both sides of PV buses' holomorphic functions.

$$\begin{aligned} &\sum_{k=1}^{N} Y_{ik}\left(V_k[0,0,\cdots 0] + V_k[1,0,\cdots,0]s_1 + V_k[0,1,\cdots,0]s_2 + \cdots\right) \\ &= \left(P_i - j\left(Q_i[0,0,\cdots 0] + Q_i[1,0,\cdots,0]s_1 + Q_i[0,1,\cdots,0]s_2 + \cdots\right)\right) \\ &\bullet\left(W_i^*[0,0,\cdots 0] + W_i^*[1,0,\cdots,0]s_1 + W_i^*[0,1,\cdots,0]s_2 + \cdots\right) \end{aligned} \tag{24}$$

The equation (24) can be reformed to a form of recursive function about $V_k[n_1,n_2,\ldots,n_D]$.

$$\begin{aligned} \sum_{k=1}^{N} Y_{ik} V_k[n_1, n_2, \cdots, n_D] &= P_i W_i^*[n_1, n_2, \cdots, n_D] - jQ_i[0,0,\cdots 0] \cdot W_i^*[n_1, n_2, \cdots, n_D] \\ &- jQ_i[n_1, n_2, \cdots, n_D] \cdot W_i^*[0,0,\cdots 0] - j \underset{k_i=1}{\overset{k_i=n_i-1}{Conv}} (Q * W_i^*)[k_1, k_2, \cdots, k_D] \end{aligned} \tag{25}$$

where $Q_i[0,0,\ldots,0]$ and $W_i[0,0,\ldots,0]$ are the obtained reactive power and the reciprocal of voltage of the physical germ solution. Note that in (25), $V_k[n_1,n_2,\ldots,n_D]$ is also dependent on the same order of extra unknowns $W^*_i[n_1,n_2,\ldots,n_D]$ and $Q_i[n_1,n_2,\ldots,n_D]$. Thus 3 additional equations need to be added to the matrix equations, by moving all the unknowns to the left hand side matrix: two for real and imaginary parts of $W^*_i[n_1,n_2,\ldots,n_D]$ and one for the real value of $Q_i[n_1,n_2,\ldots,n_D]$. For each PQ bus in the $D$-dimensional HEM, only one $D$-dimensional discrete convolution $W*V$ is induced, however, for each PV bus in the $D$-dimensional HEM, three $D$-dimensional discrete convolutions, i.e. $Q*W^*$, $W*V$ and $V*V^*$ are induced. An example of the 3-bus system of the MDHEM in Fig. 1, is introduced in detail in Appendix-B.

### *E. Transformation of Bus Type*

The MDHEM introduced above does not incorporate the reactive power limits of PV buses. Physical reactive power limits on generation units introduce discontinuities in the holomorphic functions. Let $Q_{GiMin}$ and $Q_{GiMax}$ represent the minimum and maximum reactive power limits on the generator at bus $i$, respectively. In the practice of power system power flow calculations, if the reactive power limits of generators are violated, i.e. $Q_{Gi}(s)<Q_{GiMin}$ or $Q_{Gi}(s)>Q_{GiMax}$, the bus type is changed from PV bus to PQ bus. The reactive power generation at the bus $i$ is then fixed at the value of the upper or lower limit. Since the PFEs loses its holomorphicity due to this discontinuity, the MDHEM needs to be rebuilt and resolved with altered bus types. Nevertheless, the proposed MDHEM can predict the violations of reactive power limits at PV buses in the whole solution space beforehand by evaluating $s_i$ in to $Q(s_i)$. The procedure for finding a solution of the PFE with reactive power limits on PV buses using the MDHEM is shown in Fig. 4.

### *F. Computational Resources Required by the MDHEM*

The computational resources required depend on the computation burden, the number of steps necessary to solve a problem, and memory space, the amount of storage needed while solving the problem. For $D$-dimensional HEM applied in the $N$-bus network with $s$ slack buses, $p$ PQ buses and $v$ PV buses, $N_{col}$ terms are needed to find the solution whose accuracy is up to the $M_{\text{th}}$ order, i.e. $M_{\text{th}}$-order multivariate power series. The numbers of elements for the full admittance matrix and unknown matrix on the left hand side of the $M_{\text{th}}$ order matrix equation are $(2s+2p+5v)\times(2s+2p+5v)$ and $(2s+2p+5v)\times[(M+D-1)!/M!/(D-1)!]$, respectively. The elements for the known matrix on the right hand side of the $M_{\text{th}}$ order matrix equation is also $(2s+2p+5v)\times[(M+D-1)!/M!/(D-1)!]$. The number of elements to save the $M_{\text{th}}$ order of each multivariate power series of $V(s_i)$, $W(s_i)$ and $Q(s_i)$ is

$$N_{term} = \sum_{m=0}^{M} \frac{(m+D-1)!}{m!(D-1)!}. \tag{26}$$

in which $V(s_i)$ and $W(s_i)$ are complex elements, $Q(s_i)$ is real value elements. The memory of each element is dependent on the precision digits used in the calculation, e.g. a double-precision floating-point format takes 8 bytes for each element. More memory space can be saved, if sparse calculation is used.

Most of computation burden is taken from multi-dimensional discrete convolution. There are totally $(p+3v)\times[(M+D-1)!/M!/(D-1)!]$ terms for the $M_{\text{th}}$ order matrix equation and each convolution contains $(M-1)^D-1$ multiply operations. The computation of multi-dimensional convolution can be significantly speeded up by using Row-Column Decomposition [20] or direct matrix multiplication based on Helix transform [21], [22].

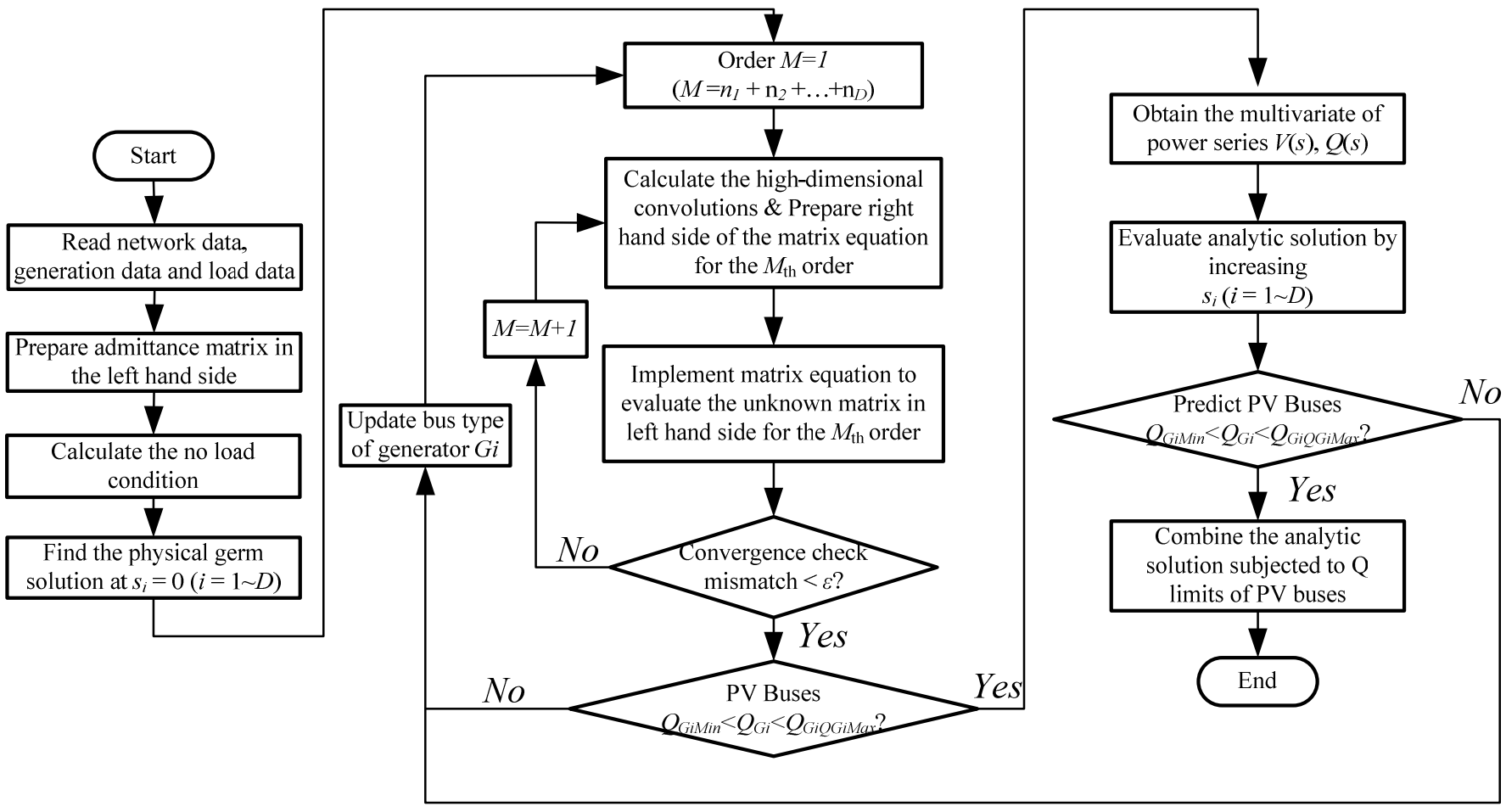

Fig. 4. The procedure of finding analytical solution of PFE using MDHEM.

## IV. Case Study

### A. Demonstration on 4-bus Power System

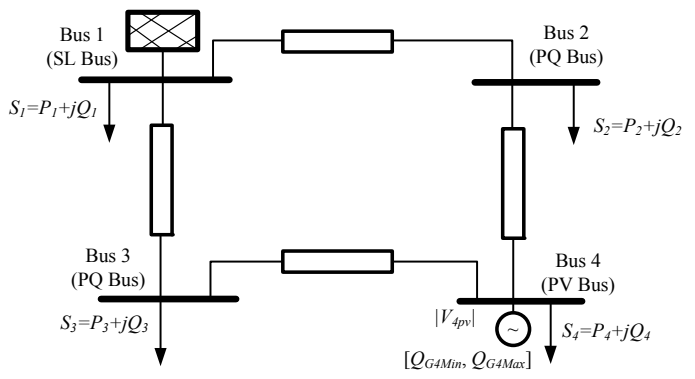

Fig. 5. One-line diagram of the 4-bus power system

As shown in Fig. 5, a 4-bus system modified from that in [19] is first used to demonstrate the MDHEM. In Case 1, Bus 1 is the slack bus with fixed voltage 1.02pu$\angle 0^\circ$, Bus 2 and Bus 3 are load buses, and Bus 4 is the generator bus maintaining its voltage magnitude at 0.98pu.

Loads at Bus 2 and Bus 3 are two independent embedding scales in the MDHEM, i.e. $s_1$ and $s_2$ controlling the loads at Bus 2 and Bus 3 respectively. The MDHEM is implemented in MATLAB using MATPOWER toolbox, which takes 1.04 sec to obtain the result with error less than $10^{-8}$ for this 2-*D* MDHEM. The result is a 12th order multivariate power series and the terms of orders up to 2 (i.e. $M$<=2) are given in the Table III.

TABLE III. Result of multivariate power series for Bus 2, 3, 4 ($M$<=2)

| Bus | V[0,0] | V[1,0] | V[1,0] | V[2,0] | V[1,1] | V[0,2] |
|---|---|---|---|---|---|---|
| $V_2$ | 9.95e-1 +6.861e-2i | -2.465e-2 -5.894e-2i | 2.248e-3 -1.833e-2i | -3.232e-3 +6.562e-6i | -1.338e-3 -1.982e-4i | -2.458e-4 -4.745e-4i |
| $V_3$ | 1.005e0+ 4.379e-2i | 1.962e-3 -1.601e-2i | -3.510e-2 -5.436e-2i | -3.138e-4 -3.872e-4i | -1.107e-3 +1.600e-4i | -3.753e-3 +2.763e-5i |
| $V_4$ | 9.727e-1 +1.197e-1i | 5.323e-3 -4.326e-2i | 3.012e-3 -3.179e-2i | -8.478e-4 -1.047e-3i | -1.277e-3 -1.291e-3i | -4.261e-4 -8.234e-4i |

Fig. 6 shows the results of bus voltages by evaluating the multivariate power series obtained by the MDHEM, compared with the results of the Newton-Raphson (N-R) method calculated for every operating condition with granularity of $s_i$=0.1. Fig. 7 shows the difference between the results from the N-R and MDHEM. Note that if the N-R does not converge, the result is assigned to 0 here for the sake of clarity and the non-convergence of N-R does not theoretically signify the non-existence of solutions. It can be observed that in the N-R's convergence region, the solutions from the MDHEM show great consistency, (maximum error $3.454\times10^{-4}$pu at Bus 3).

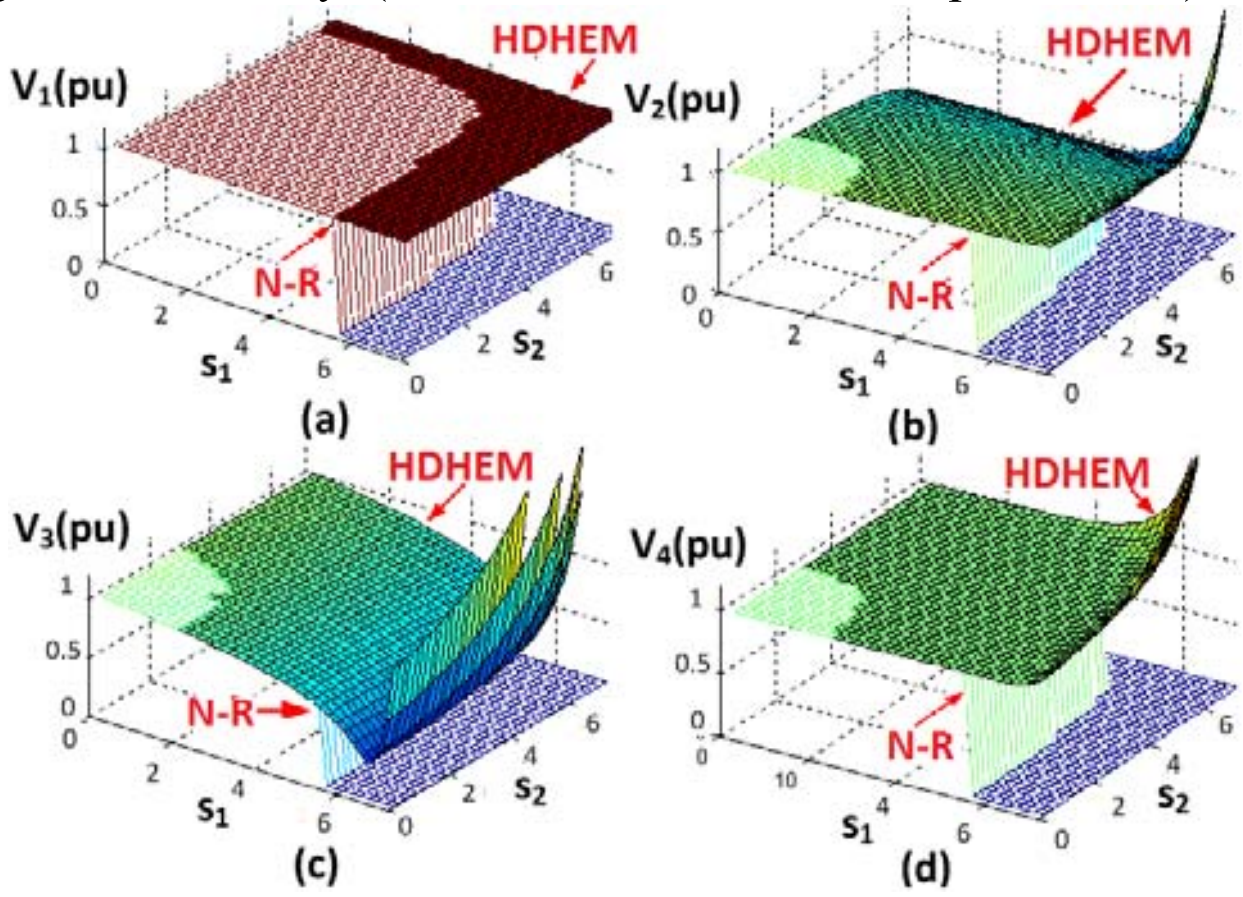

Fig. 6. Comparison of results with PV bus between N-R and MDHEM.

In the N-R's non-convergence region, there is some mismatch. The results directly from multivariate power series are not able to clearly signify the voltage stability boundary. However, that may be improved by using multivariate Pade approximants and will be our future research.

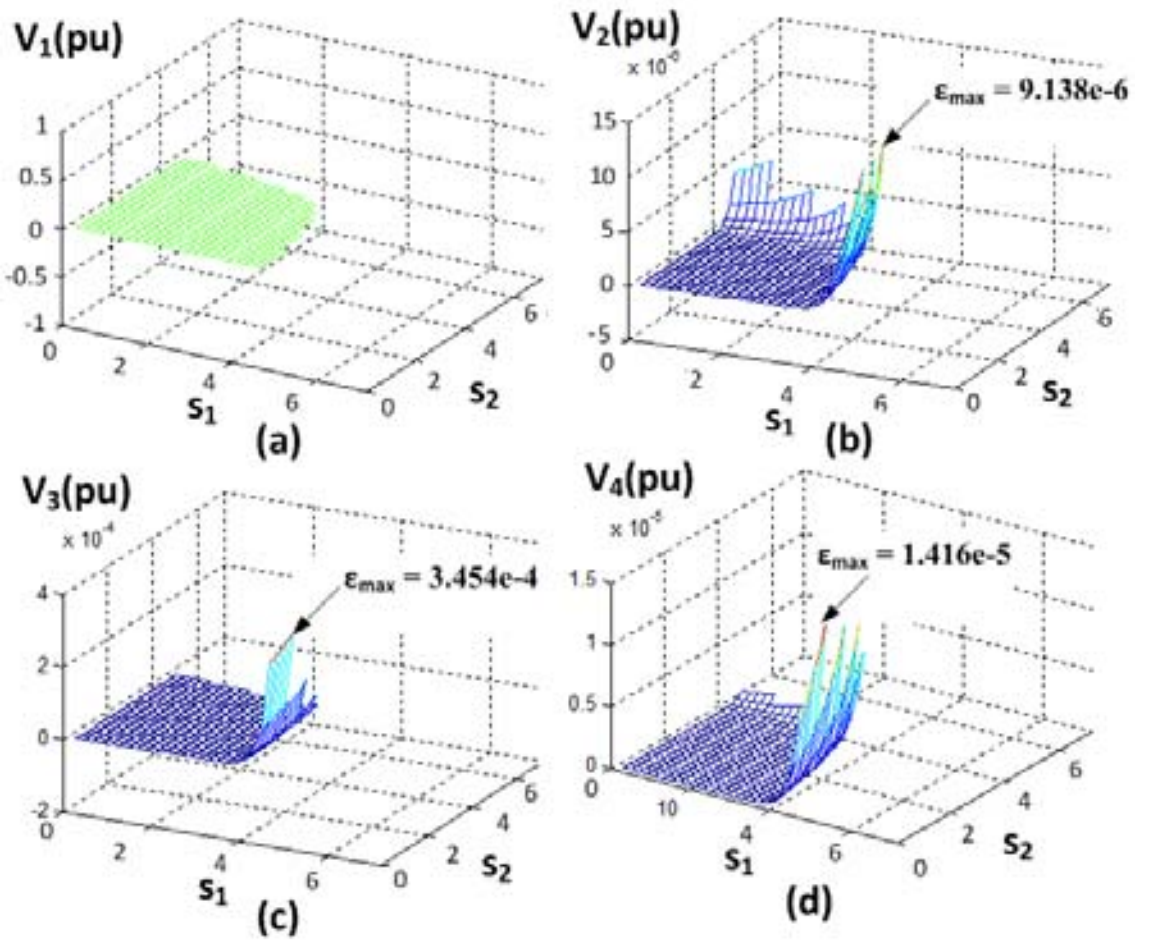

Fig. 7. Difference between MDHEM and N-R in Case 1 with PV bus.

In Case 2, with the increase of loading scales at load buses, the reactive power injection at the generator bus (Bus 2) violates its upper limit of 100MVar. The PV bus will be transformed to a fixed PQ bus. Fig. 8 shows the results of bus voltages compared with the results of the N-R method for Case 2 and Fig. 9 shows the difference between the MDHEM and N-R. It can be observed that $V_4$ is no longer a PV bus with fixed voltage magnitude and the convergence region is smaller than that in Case 1. The maximum error is larger, (i.e. 0.0529pu at Bus 2), because higher nonlinearity is induced, especially for the heavily loaded conditions on the surface.

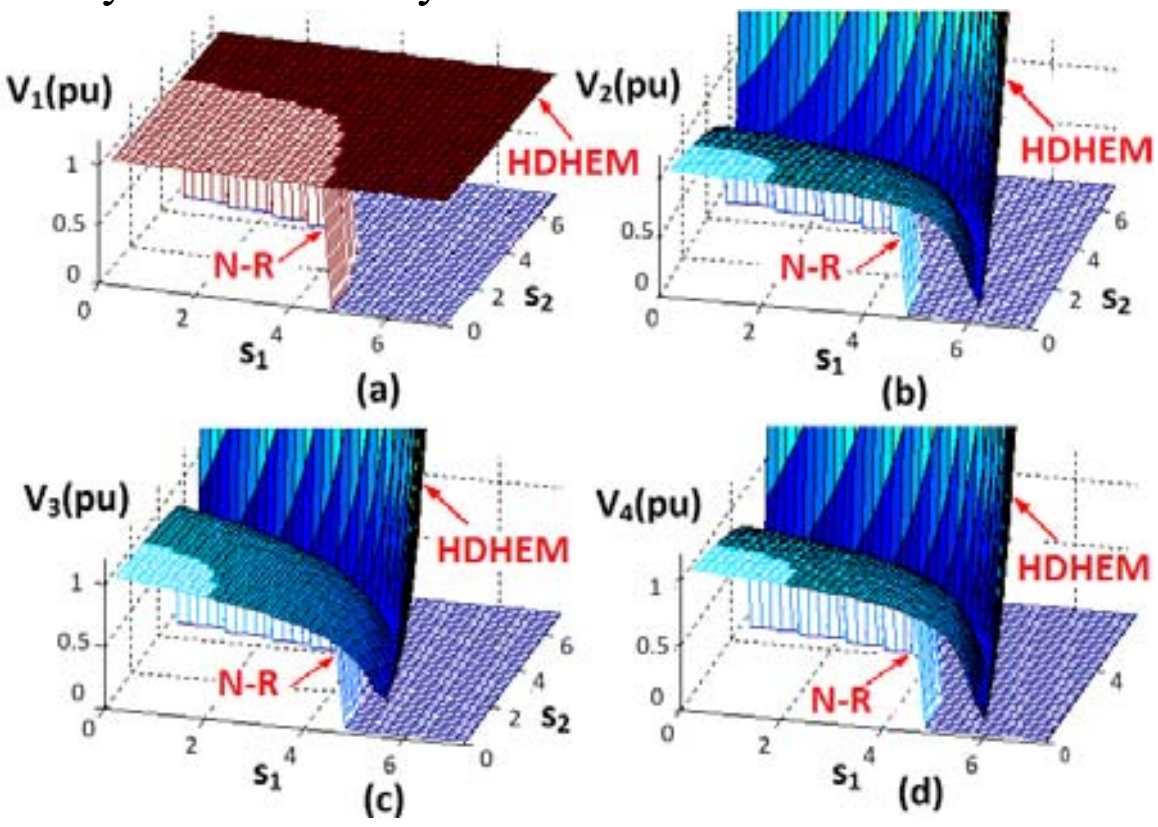

Fig. 8. Comparison of results without PV bus between N-R and MDHEM.

Fig 10 shows the comparison of reactive power and voltage magnitude for Case 1 and Case 2. It can be observed in Fig. 10(a) that the reactive power for the PV bus is a multivariate nonlinear function. The section of the two surfaces is the Q limit boundary with respect to different loading scales. In Fig. 10(b), inside the Q limit boundary, the voltage of PV bus is higher than that in the case of PQ bus and vice versa. The Q limit boundary in the high-dimensional space can be clearly predicted by solutions from the MDHEM.

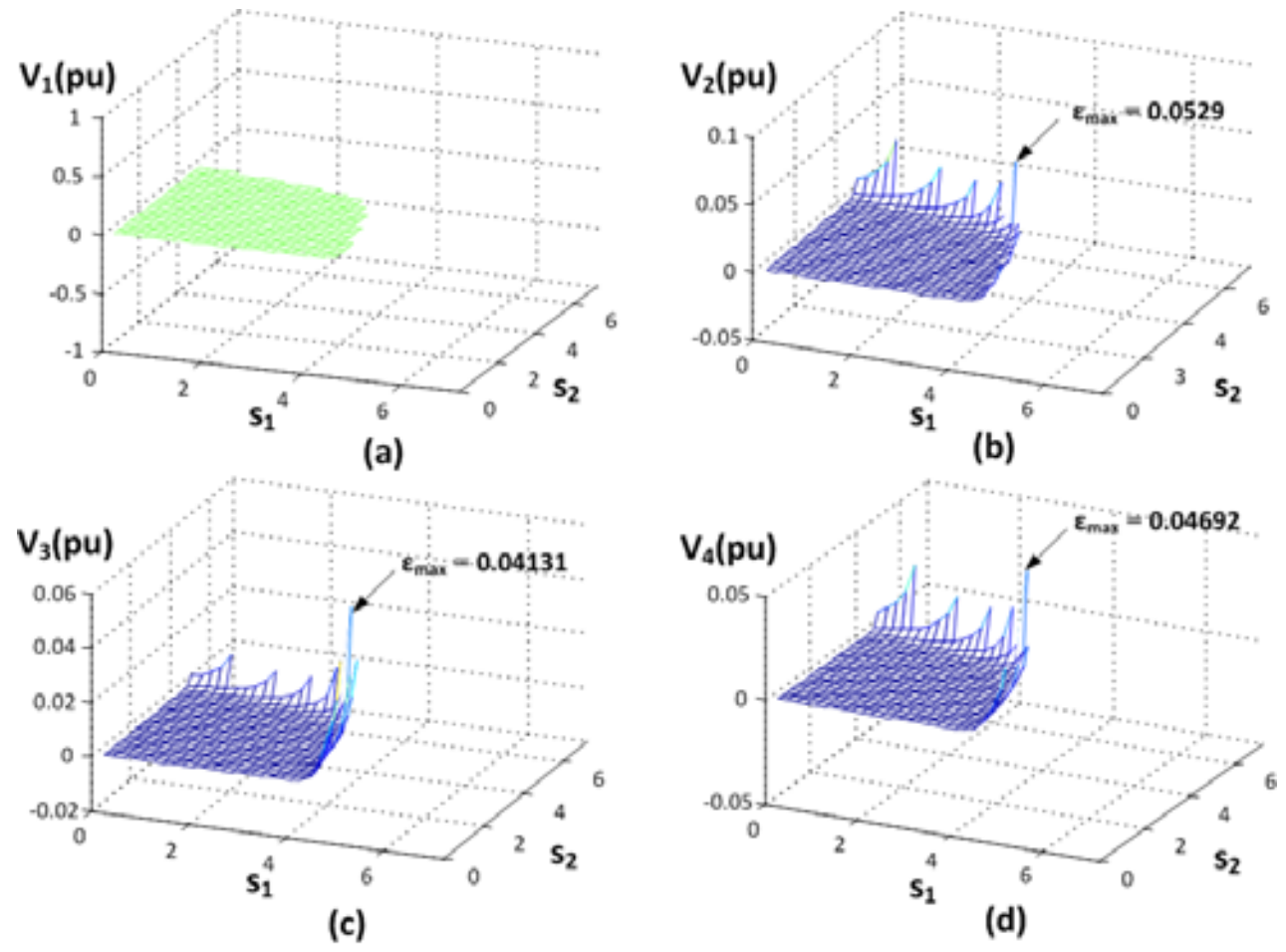

Fig. 9. Difference between MDHEM and N-R Case 2 without PV bus.

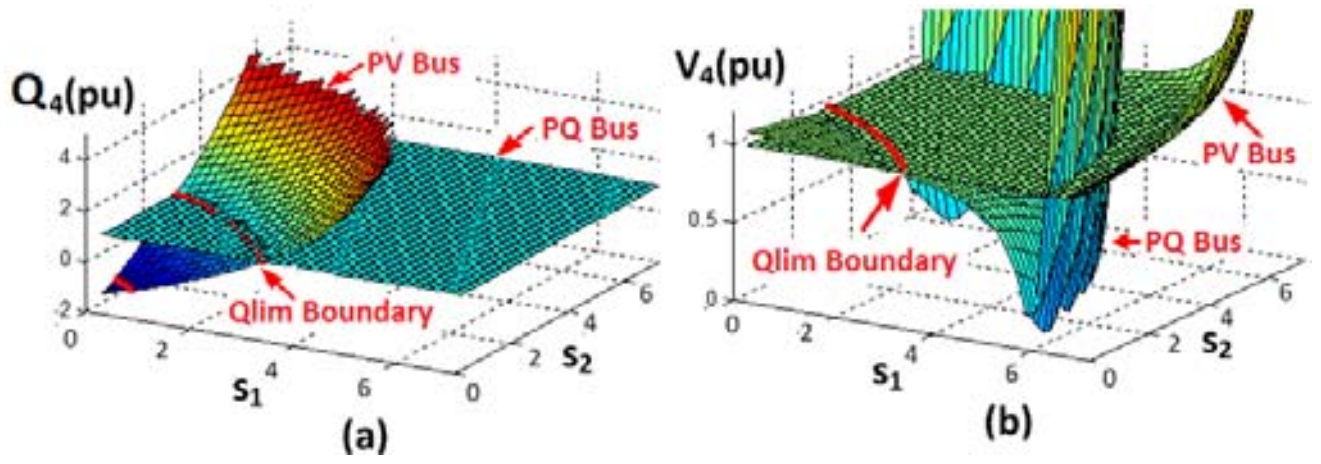

Fig. 10. The (a) reactive power and (b) voltage of PV bus for Case 1 and 2.

## B. Demonstration on 14-bus Power System

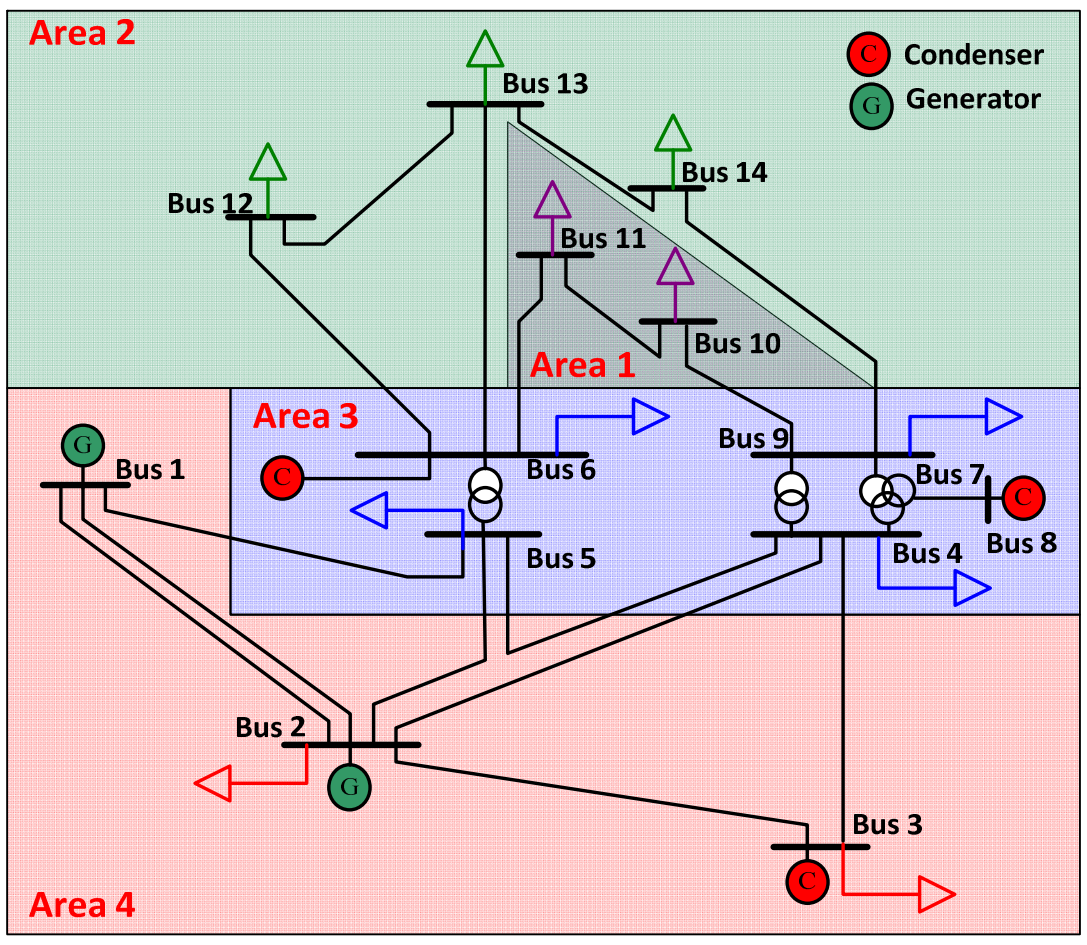

Fig. 11. One-line diagram of the demonstrative 14-bus system.

The MDHEM is also demonstrated on the IEEE 14-bus test system. The loads are geographically grouped to 4 areas with their respective loading scales $s_1$, $s_2$, $s_3$ and $s_4$, as shown in Fig. 11. The 4-dimensional MDHEM is performed and the result of a 11th order 4-*D* multivariate power series is obtained in 12.56sec in MATLAB. The error tolerance is also $1\times10^{-8}$ pu.

A load bus, i.e. Bus 5, is randomly selected to observe the voltage with respect to the four different scales, as shown in Fig 12. Keeping the load level of Area 4 unchanged ($s_4$=1), the voltages of load buses decrease with the increase of loading scales, i.e. the X-axis and Y-axis for $s_1$ and $s_2$ respectively, and the different surface of $s_3$.

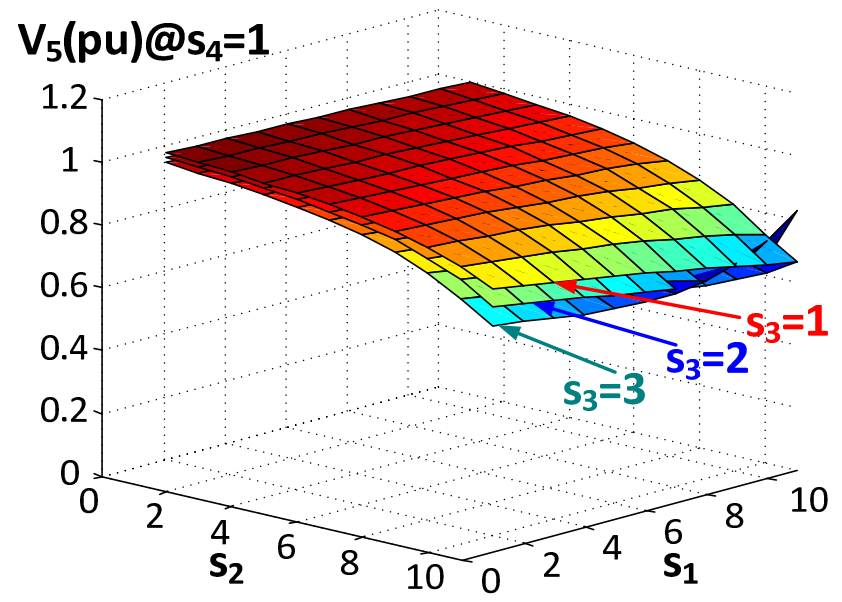

Fig. 12. Voltage of Bus 5 with respect to other scales for $s_4$= 1.

It can also be concluded that although the computation speed is lower with higher dimensions, the convergence rate does not dependent on the embedding dimensions. This makes maximum order of multivariate power series to be manageable.

## V. Conclusion and Discussion

This paper proposed a new multi-dimensional holomorphic embedding method (MDHEM) for solving the power flow equations and the explicit solutions are obtained for all operating conditions in the high-dimensional solution space. The voltage vector and power of each bus can be explicitly expressed by a convergent multivariate power series of all the loads. Compared with the traditional iterative methods for power flow calculations and inaccurate sensitivity analysis method for voltage control, the MDHEM can prepare the algebraic variables of a power system in all operating conditions offline and evaluate them online by only plugging in the values of the required operating conditions into the scales of the non-linear multivariate power series. The result of the MDHEM can also predict the reactive power limits for perspective operating conditions in advance, giving operators with enough time to take proactive actions. This method not only provides a tool to obtain the explicit power-flow solutions of any power systems, but may also to explore the nonlinearity of power flow equations.

This new method has been demonstrated on the 4-bus power system and the 14-bus IEEE standard power systems. The error is acceptable and may be further reduced by using multivariate Pade approximants.

## Appendix

### A. Example-A: Finding the physical germ solution in 3-bus system

A 3-bus system, shown in Fig. 1, is adopted to demonstrate the procedure of finding the physical germ solution. The first step is to calculate the initial no-load no-generation condition (Point *A* in Fig. 2). Only the slack bus propagates its voltage to the whole passive network.

$$\begin{bmatrix} 1 & & & & & \\ & 1 & & & & \\ G_{21} & -B_{21} & G_{22} & -B_{22} & G_{23} & -B_{23} \\ B_{21} & G_{21} & B_{22} & G_{22} & B_{23} & G_{23} \\ G_{31} & -B_{31} & G_{32} & -B_{32} & G_{33} & -B_{33} \\ B_{31} & G_{31} & B_{32} & G_{32} & B_{33} & -B_{33} \end{bmatrix} \begin{bmatrix} V_{g1re}[0] \\ V_{g1im}[0] \\ V_{g2re}[0] \\ V_{g2im}[0] \\ V_{g3re}[0] \\ V_{g3im}[0] \end{bmatrix} = \begin{bmatrix} \mathrm{Re}\left(V_1^{SL}\right) \\ \mathrm{Im}\left(V_1^{SL}\right) \\ 0 \\ 0 \\ 0 \\ 0 \end{bmatrix} \quad \text{(A1)}$$

The second step is to find the physical germ solution by a simple embedding, i.e. 3rd column in TABLE I. The right hand side of the PQ bus (i.e. Bus 3) is 0, since no-load condition is held for the physical germ solution. Thus, the embedding of PV bus (i.e. Bus 2) is needed to gradually adjust its voltage magnitude to the specified value $|V_{2pv}|$, while keep the active power generation as the specified value $P_{G2}$.

$$\begin{cases} \sum_{k=1}^{N} Y_{ik} V_{gi}[n] = P_i W_i^*[n] - jQ_{gi}[n] W_i^*[0] - j\left(\sum_{k=1}^{n-1} Q_{gi}[k] W_i^*[n-k]\right) \\ V_{gi}[n] V_{gi}^*[0] + V_{gi}[n] V_{gi}^*[0] + \overset{n-1}{\underset{1}{Conv}}(V * V^*) = \delta_{n0} \cdot 2\left|V_{gi}[0]\right|^2 + \delta_{n1} \cdot \left(\left|V_i^{sp}\right|^2 - \left|V_{gi}[0]\right|^2\right) \end{cases} \tag{A2}$$

where $\delta_{ni}$ is Kronecker delta function that equals to 1 only for order of $i = n$ and vanish for the other orders.

$$\delta_{ni} = \begin{cases} 1 & \text{if} \quad n = i \\ 0 & \text{otherwise} \end{cases} \tag{A3}$$

Separate the real and imaginary parts of the matrix equation and put all the $n^{\text{th}}$ order terms (i.e. the unknowns) to the left hand side and leave terms 0 to ($n$-1) (i.e. the knowns) the right hand side. The matrix equation is extended by adding $W_2(s)$ and $Q_2(s)$ to the left hand side. The error of physical germ in PFE will quickly converge to 0 just in several recursions, since the deviation of voltage at PV Bus 2 contains high order terms of $V_2(s)$.

$$\begin{bmatrix} 1 & 0 & 0 & 0 & 0 & 0 & 0 & 0 & 0 \\ 0 & 1 & 0 & 0 & 0 & 0 & 0 & 0 & 0 \\ G_{21} & -B_{21} & G_{22} & -B_{22} & G_{23} & -B_{23} & 0 & 0 & W_{g2im}[0] \\ B_{21} & G_{21} & B_{22} & G_{22} & B_{23} & G_{23} & 0 & 0 & W_{g2re}[0] \\ G_{31} & -B_{31} & G_{32} & -B_{32} & G_{33} & -B_{33} & 0 & 0 & 0 \\ B_{31} & G_{31} & B_{32} & G_{32} & B_{33} & -G_{33} & 0 & 0 & 0 \\ 0 & 0 & W_{g2re}[0] & -W_{g2im}[0] & 0 & 0 & V_{g2re}[0] & -V_{g2im}[0] & 0 \\ 0 & 0 & W_{g2im}[0] & W_{g2re}[0] & 0 & 0 & V_{g2im}[0] & V_{g2re}[0] & 0 \\ 0 & 0 & V_{g2re}[0] & V_{g2im}[0] & 0 & 0 & 0 & 0 & 0 \end{bmatrix} \begin{bmatrix} V_{g1re}[n] \\ V_{g1im}[n] \\ V_{g2re}[n] \\ V_{g2im}[n] \\ V_{g3re}[n] \\ V_{g3im}[n] \\ W_{g2re}[n] \\ W_{g2im}[n] \\ Q_{g2}[n] \end{bmatrix} = \begin{bmatrix} 0 \\ 0 \\ \mathrm{Re}\left(P_2 W_{g2re}[n-1] + \overset{n-1}{\underset{1}{Conv}}(Q * W^*)\right) \\ \mathrm{Im}\left(P_2 W_{g2re}[n-1] + \overset{n-1}{\underset{1}{Conv}}(Q * W^*)\right) \\ 0 \\ 0 \\ -\mathrm{Re}\left(\overset{n-1}{\underset{1}{Conv}}(W * V)\right) \\ -\mathrm{Im}\left(\overset{n-1}{\underset{1}{Conv}}(W * V)\right) \\ \delta_{n1} \cdot \varepsilon[1] - \frac{1}{2}\overset{n-1}{\underset{1}{Conv}}(V * V^*) \end{bmatrix} \tag{A4}$$

in which

$$\varepsilon[1] = \frac{1}{2}\left(\left|V_i^{sp}\right|^2 - \left|V_{gi}[0]\right|^2\right) \tag{A5}$$

## B. Example-B: the MDHEM in 3-bus system

Assume $s_1$ and $s_2$ scale the active and reactive power of the PQ bus respectively. The matrix equation for $M$th order is shown correspondingly in (A7), where the column sequence of the unknowns' matrix and the knowns' matrix are 2-polytope numbers, which holds

$$M = n_1 + n_2 \,. \tag{A6}$$

$$\begin{bmatrix} 1 & 0 & 0 & 0 & 0 & 0 & 0 & 0 & 0 \\ 0 & 1 & 0 & 0 & 0 & 0 & 0 & 0 & 0 \\ G_{21} & -B_{21} & G_{22} & -B_{22} & G_{23} & -B_{23} & -P_{2g} & Q_{2g} & W_{g2im} \\ B_{21} & G_{21} & B_{22} & G_{22} & B_{23} & G_{23} & Q_{2g} & P_{2g} & W_{g2re} \\ G_{31} & -B_{31} & G_{32} & -B_{32} & G_{33} & -B_{33} & 0 & 0 & 0 \\ B_{31} & G_{31} & B_{32} & G_{32} & B_{33} & -G_{33} & 0 & 0 & 0 \\ 0 & 0 & W_{g2re} & -W_{g2im} & 0 & 0 & V_{g2re} & -V_{g2im} & 0 \\ 0 & 0 & W_{g2im} & W_{g2re} & 0 & 0 & V_{g2im} & V_{g2re} & 0 \\ 0 & 0 & V_{g2re} & V_{g2im} & 0 & 0 & 0 & 0 & 0 \end{bmatrix} \begin{bmatrix} V_{1re}[M,0] & V_{1re}[M-1,1] & \cdots & V_{1re}[n_1,n_2] & \cdots & V_{1re}[0,M] \\ V_{1im}[M,0] & V_{1im}[M-1,1] & \cdots & V_{1im}[n_1,n_2] & \cdots & V_{1im}[0,M] \\ V_{2re}[M,0] & V_{2re}[M-1,1] & \cdots & V_{2re}[n_1,n_2] & \cdots & V_{2re}[0,M] \\ V_{2im}[M,0] & V_{2im}[M-1,1] & \cdots & V_{2im}[n_1,n_2] & \cdots & V_{2im}[0,M] \\ V_{3re}[M,0] & V_{3re}[M-1,1] & \cdots & V_{3re}[n_1,n_2] & \cdots & V_{3re}[0,M] \\ V_{3im}[M,0] & V_{3im}[M-1,1] & \cdots & V_{3im}[n_1,n_2] & \cdots & V_{3im}[0,M] \\ W_{2re}[M,0] & W_{2re}[M-1,1] & \cdots & W_{2re}[n_1,n_2] & \cdots & W_{2re}[0,M] \\ W_{2im}[M,0] & W_{2im}[M-1,1] & \cdots & W_{2im}[n_1,n_2] & \cdots & W_{2im}[0,M] \\ Q_2[M,0] & Q_2[M-1,1] & \cdots & Q_2[n_1,n_2] & \cdots & Q_2[0,M] \end{bmatrix} = \begin{bmatrix} 0 & \cdots & 0 & \cdots & 0 \\ 0 & \cdots & 0 & \cdots & 0 \\ \mathrm{Im}\left(\overset{M-1}{\underset{k_1=1}{Conv}}(Q*W^*)\right) & \cdots & \mathrm{Im}\left(\overset{n_1-1,n_2-1}{\underset{k_1=1,k_2=1}{Conv}}(Q*W^*)\right) & \cdots & \mathrm{Im}\left(\overset{M-1}{\underset{k_1=1}{Conv}}(Q*W^*)\right) \\ -\mathrm{Re}\left(\overset{M-1}{\underset{k_1=1}{Conv}}(Q*W^*)\right) & \cdots & -\mathrm{Re}\left(\overset{n_1-1,n_2-1}{\underset{k_1=1,k_2=1}{Conv}}(Q*W^*)\right) & \cdots & -\mathrm{Re}\left(\overset{M-1}{\underset{k_1=1}{Conv}}(Q*W^*)\right) \\ \mathrm{Re}\left(P_3 W_3^*[M-1,0]\right) & \cdots & \mathrm{Re}\left(P_3 W_3^*[n_1-1,n_2] - jQ_3 W_3^*[n_1,n_2-1]\right) & \cdots & \mathrm{Re}\left(-jQ_3 W_3^*[0,M-1]\right) \\ \mathrm{Im}\left(P_3 W_3^*[M-1,0]\right) & \cdots & \mathrm{Im}\left(P_3 W_3^*[n_1-1,n_2] - jQ_3 W_3^*[n_1,n_2-1]\right) & \cdots & \mathrm{Im}\left(-jQ_3 W_3^*[0,M-1]\right) \\ -\mathrm{Re}\left(\overset{M-1}{\underset{k_1=1}{Conv}}(W*V)\right) & \cdots & -\mathrm{Re}\left(\overset{n_1-1,n_2-1}{\underset{k_1=1,k_2=1}{Conv}}(W*V)\right) & \cdots & -\mathrm{Re}\left(\overset{M-1}{\underset{k_2=1}{Conv}}(W*V)\right) \\ -\mathrm{Im}\left(\overset{M-1}{\underset{k_1=1}{Conv}}(W*V)\right) & \cdots & -\mathrm{Im}\left(\overset{n_1-1,n_2-1}{\underset{k_1=1,k_2=1}{Conv}}(W*V)\right) & \cdots & -\mathrm{Im}\left(\overset{M-1}{\underset{k_2=1}{Conv}}(W*V)\right) \\ -\frac{1}{2}\overset{M-1}{\underset{k_1=1}{Conv}}(V*V^*) & \cdots & -\frac{1}{2}\overset{n_1-1,n_2-1}{\underset{k_1=1,k_2=1}{Conv}}(V*V^*) & \cdots & -\frac{1}{2}\overset{M-1}{\underset{k_2=1}{Conv}}(V*V^*) \end{bmatrix} \tag{A7}$$